# Comprehensive characterization of tumor therapeutic response with simultaneous mapping cell size, density, and transcytolemmal water exchange

Diwei Shi[1], Sisi Li[2], Fan Liu[2], Xiaoyu Jiang[3,4], Lei Wu[5], Li Chen[1], Quanshui Zheng[1], Haihua Bao[5], Hua Guo[2], Junzhong Xu[3,4,6,7*]

1  Center for Nano and Micro Mechanics, Department of Engineering Mechanics, Tsinghua University, Beijing 100084, China
2  Center for Biomedical Imaging Research, Department of Biomedical Engineering, School of Medicine, Tsinghua University, Beijing 100084, China
3  Institute of Imaging Science, Vanderbilt University Medical Center, Nashville, TN 37232, USA
4  Department of Radiology and Radiological Sciences, Vanderbilt University Medical Center, Nashville, TN 37232, USA
5  Qinghai University Affiliated Hospital, Qinghai, Xining 810000, China
6  Department of Biomedical Engineering, Vanderbilt University, Nashville, TN 37232, USA
7  Department of Physics and Astronomy, Vanderbilt University, Nashville, TN 37232, USA

\* **Corresponding author**: Address: Vanderbilt University, Institute of Imaging Science, 1161 21st Avenue South, AA 1105 MCN, Nashville, TN 37232-2310, United States. Fax: +1 615 322 0734.  E-mail address: junzhong.xu@vumc.org (Junzhong Xu)

## ABSTRACT

Early assessment of tumor therapeutic response is an important topic in precision medicine to optimize personalized treatment regimens and reduce unnecessary toxicity, cost, and delay. Although diffusion MRI (dMRI) has shown potential to address this need, its predictive accuracy is limited, likely due to its unspecific sensitivity to overall pathological changes. In this work, we propose a new quantitative dMRI-based method dubbed EXCHANGE (MRI of water Exchange, Confined and Hindered diffusion under Arbitrary Gradient waveform Encodings) for simultaneous mapping of cell size, cell density, and transcytolemmal water exchange. Such rich microstructural information comprehensively evaluates tumor pathologies at the cellular level. Validations using numerical simulations and in vitro cell experiments confirmed that the EXCHANGE method can accurately estimate mean cell size, density, and water exchange rate constants. The results from in vivo animal experiments show the potential of EXCHANGE for monitoring tumor treatment response. Finally, the EXCHANGE method was implemented in breast cancer patients with neoadjuvant chemotherapy, demonstrating its feasibility in assessing tumor therapeutic response in clinics. In summary, a new, quantitative dMRI-based EXCHANGE method was proposed to comprehensively characterize tumor microstructural properties at the cellular level, suggesting a unique means to monitor tumor treatment response in clinical practice.

# 1 INTRODUCTION

With the rapid development of more effective treatments, imaging plays a more important role not only in diagnosis but also in assisting therapy. For example, neoadjuvant (pre-operative) chemotherapy (NAC) is one of the major treatment options for operable early-stage breast cancer (1). The assessment of tumor early response to NAC may have a significant impact on patient-specific treatment strategy. Earlier identification of drug resistance can prompt the discontinuation of ineffective therapy (avoiding unnecessary toxicity), and switch to a potentially more effective alternative method (avoiding treatment delays). However, current imaging criteria for evaluating therapeutic response are based on Response Evaluation Criteria in Solid Tumors (RECIST) guidelines (2), which are solely based on tumor shrinkage, a *downstream* effect of various treatment-induced molecular and cellular changes. Therefore, RECIST criteria are frequently inadequate for assessing early therapeutic response. Currently, there are numerous potentially useful imaging techniques to address this clinical need, including dynamic contrast-enhanced (DCE) (3,4) and diffusion-weighted MRI (DWI) (5,6). The DWI-derived apparent diffusion coefficient (ADC) has been shown to provide additional and complementary information about tissue cellularity (7) and microstructure that can be used to characterize breast tumors and to monitor their response to treatment (6).

However, clinical trials found ADC only demonstrated moderate prediction performance of treatment response, particularly predicting pathological complete response (pCR) in breast cancer that is associated with long-term outcomes and is a potential surrogate marker for survival (4). One explanation is that ADC represents averaged diffusion that is influenced by all pathological changes, including those with competing effects (8). To enhance the specificity to specific pathological changes and improve the predictive performance of treatment response, multiple diffusion MRI (dMRI) based methods have been developed, such as the single-compartmental, time-dependent ADC (8,9) and multi-compartmental biophysical models (10-12) to provide detailed pathophysiological information. These methods, such as IMPULSED (13,14) and VERDICT (15), use both multiple b values and a broad range of diffusion times, and establish multi-compartmental biophysical models to extract microstructural information at the cellular level, such as cell size (effective diameter $d$), intracellular volume fraction ($v_{in}$), intra- and extracellular diffusivities ($D_{in}$ and $D_{ex}$). Previous studies have demonstrated the clinical potential, particularly in breast cancer (16,17) and prostate cancer (18).

However, the current multi-compartment models usually ignore transcytolemmal water exchange (i.e., the water exchange across cell membranes). This is because the entanglement of water exchange and diffusion makes it challenging for biophysical modeling, particularly for finite duration of gradients. However, water exchange causes a mixture of intra- and extracellular water molecules. It is theoretically inappropriate to separate dMRI signals and assume that they are arising from multiple independent compartments. In such cases, the ignorance of water exchange could result in significant bias in estimating microstructural parameters. For example, it has been found that although cancer cell diameter can still be accurately estimated, the intracellular volume fraction $v_{in}$ is significantly underestimated due to non-negligible water exchange(19). This prevents these quantitative multi-compartment dMRI methods from providing accurate information on cell density and is unable to estimate cell membrane permeability (20,21).

Meanwhile, there is another type of dMRI method that focuses on transcytolemmal water exchange, including the Kärger model (22,23) and many variant methods such as FEXI (24), NEXI (25), SMEX(26), and diffusion time-dependent kurtosis imaging (27,28). The Kärger model describes the magnetizations of two Gaussian compartments undergoing water exchange, providing an opportunity to quantify transcytolemmal water exchange rate constants without using contrast agents as in DCE MRI (29,30). Note that cross-membrane water exchange is associated with tumor malignancy (29) and metabolic activity of cells (31), which has been shown as an important indicator of disease status at the cellular level (32-34).

It is of great interest to perform a multi-parametric MRI to provide a comprehensive characterization of tumors including cell size, cell density, and transcytolemmal water exchange. Unfortunately, this would significantly increase the total scan time which is not desirable in clinical practice. An ideal approach is to incorporate transcytolemmal water exchange in the multi-compartmental model. This can not only provide comprehensive microstructural information but also enhance the accuracy of estimations of cell size and density (35,36). Some limited efforts have tried to incorporate the impact of intracellular restricted diffusion into the Kärger model (37-40), such as the JOINT model (36), enabling simultaneous estimation of cell sizes and water exchange rate constants. However, these Kärger model-based methods are only valid with relatively slow water exchange, and their performance was still unreliable in the case of fast exchange (41). It is plausible to develop a fast and accurate MRI method that provides comprehensive microstructural information including cell size, density, and transcytolemmal water exchange.

In this work, we propose a new dMRI-based microstructural method to address the above needs. Specifically, this method uses an integrated biophysical model that combines both an extended Kärger model for arbitrary gradient waveforms and the IMPULSED framework. A novel two-mode diffusion model is also introduced to quantify the influence of transcytolemmal water exchange on intracellular diffusion and a practical correction for edge-enhancement effect is proposed to provide more accurate estimations of exchange rate constants. We name it the EXCHANGE method since it includes water Exchange, Confined (restricted), and Hindered diffusion under Arbitrary Gradient waveform Encodings. Comprehensive validations, including numerical simulations and retrospective in vitro cell experiments, were performed. In retrospective in vivo animal experiments and a proof-of-concept study, we demonstrated the clinical potential of the EXCHANGE method, where the total scan time is ~6 minutes in monitoring tumor therapeutic response in breast cancer patients with neoadjuvant chemotherapy. The open-sourced sequence and data analysis code will make the EXCHANGE method readily achievable in clinical trials.

## 2  MATERIALS AND METHODS

### 2.1  Theory of the EXCHANGE method

Kärger model-based methods are only valid with relatively slow water exchange, and their performance was unreliable for fast exchange (41). Generally, these methods have the following limitations:

1) Retain the traditional short pulse approximation and ignore water exchange when diffusion gradients are on, which is inappropriate for diffusion sequences with long gradient durations, such as OGSE (42) and q-vector trajectory (43).
2) Neglect the fact that intracellular diffusion contains both restricted diffusion (confined inside cells) and hindered diffusion (cross membranes), but still use restricted intracellular diffusion only, leading to biases in model fitting.
3) Assume that the magnetization exchange rate constants between the intra- and extracellular compartments ($k_{in}^m$ and $k_{ex}^m$) are equal to the true water molecule exchange rate constants $k_{in}$ and $k_{ex}$. In fact, the restriction-induced edge-enhancement effect always results in $k_{in}^m > k_{in}$ (44), i.e., overestimation of exchange rate constant (45).

 Here we propose a novel biophysical model to address all the above limitations. Detailed procedures are listed below and summarized in Figure 1. Briefly, assuming dMRI signals arise from the intracellular and extracellular compartments, we extended and revised the Kärger model as:

$$\begin{aligned}\frac{dM_{in}}{dt} &= -\frac{bD_{in}^*}{T}M_{in} - k_{in}^m M_{in} + k_{ex}^m M_{ex} \\ \frac{dM_{ex}}{dt} &= -\frac{bD_{ex}^*}{T}M_{ex} - k_{ex}^m M_{ex} + k_{in}^m M_{in}\end{aligned} \quad (1)$$

where $M_{in}$ and $M_{ex}$ are the intra- and extracellular magnetizations. $D_{in}^*$ and $D_{ex}^*$ are the modified diffusivities within the compartments, $k_{in}^m$ and $k_{ex}^m$ are the exchange rate constants of magnetizations (from 'in' to 'ex' and from 'ex' to 'in'), and $T$ equals the sum of the gradient duration $\delta$ and separation $\Delta$ for any diffusion sequences.

By solving the differential Eq. (1), the normalized diffusion-weighted signal $S$ can be expressed as the following linear combinations of exponential terms:

$$S = \frac{M_{in} + M_{ex}}{M_0} = V_1 \exp(-bD_1^*) + (1 - V_1)\exp(-bD_2^*) \quad (2)$$

where $M_0$ is the original non-diffusion-weighted magnetization, $D_1^*$, $D_2^*$, and $V_1$ are the parameters related to the unknown quantities $D_{in}^*$, $D_{ex}^*$, $k_{in}^m$, and $k_{in}^m$ in Eq. (1), which contains the microstructural information. The theoretical details are summarized in the following subsections and **Supplemental Materials**.

2.1.1 Two-mode intracellular diffusion

Most previous studies have defined intra- and extracellular diffusion as restricted (i.e., confined inside impermeable cells) and hindered, respectively, as shown in the IMPULSED and VERDICT models. In the presence of transcytolemmal water exchange, it is intuitive to replace the Gaussian diffusivities in the Kärger model with the apparent restricted diffusion coefficient $ADC_r$ and the hindered diffusivity $D_{ex}$, which has been implemented in the JOINT model (36). However, this simplification will result in biases when fitting microstructural parameters. Apparently, due to the presence of water exchange, the actual intracellular diffusion includes not only restricted diffusion but also hindered diffusion caused by water molecules crossing membranes. We hereby propose a heuristic two-mode intracellular diffusion model to describe this fact as shown in Figure 1a.

We defined the probability of an intracellular molecule crossing the membrane and then moving to the extracellular space as $p$, and estimated it from the existing parameters, including cell radius $R$, intracellular diffusivity $D_{in}$ and water exchange rate constant $k_{in}$:

$$p = \frac{\bar{t}^*}{\bar{t}^* + 1/k_{in} - \bar{t}} \quad (3)$$

where $\bar{t}$ and $\bar{t}^*$ equal to $(3R/4)^2/(2D_{in})$ and $(4R/3)^2/(2D_{in})$, respectively. The relevant details are shown in the **Supplemental Materials**.

Then, the diffusion of intracellular water molecules was divided into two modes:

**Mode A:** For molecules that stay inside the cell, i.e., restricted diffusion, $ADC_r$ is used to describe the intensity of the diffusion movement.

**Mode B:** For molecules that cross the membrane, leave the cell, and undergo hindered diffusion, an average hindered diffusivity $D_{inh}$ is introduced and we approximated it as a linear combination of $ADC_r$ and $D_{ex}$ (the commonly used extracellular diffusivity), with the volume fractions $v_{in}$ and $v_{ex}$ as the weights, i.e., $D_{inh} = v_{in} ADC_r + v_{ex} D_{ex}$.

Finally, the two-mode diffusion coefficient of the intracellular compartment $D_{in}^*$ in Eq. (1)) can be calculated by the following approximation:

$$D_{in}^* = -\ln\big((1-p)\exp(-b \cdot ADC_r) + p\exp(-b \cdot D_{inh})\big)/b \tag{4}$$

Theoretically, extracellular diffusion also includes two modes. But considering the narrow, interstitial extracellular space, we model them both as hindered diffusion for simplicity and use an effective diffusivity $D_{ex}^*$ in Eq. (1) to characterize the overall extracellular diffusion and: $D_{ex}^* \approx D_{ex}$. Figure 1 (a) graphically illustrates the above-mentioned diffusion movements of water molecules in the intra- and extracellular spaces.

2.1.2   Correction for the edge-enhancement effect

The non-negligible impact of the restriction-induced edge-enhancement effect, as shown in our previous work (45) and Figure 1(b), is also included in the proposed model. The actual $k_{in}^m$ and $k_{ex}^m$ are usually unequal to the exchange rate constants of water molecules, i.e., ($k_{in}$ and $k_{ex}$), and typically $k_{in}^m > k_{in}$. However, it is challenging to obtain an analytical expression of $k_{in}^m$ or $k_{ex}^m$. Here, we construct an approximate form for $k_{in}^m$ based on the dimensional analysis:

$$k_{in}^m = k_{in} \cdot \left(1 + \alpha \cdot \left(\frac{bd^2}{\tau_{in}}\right)^{\gamma_1} \cdot \left(\frac{d^2}{D_{in}\tau_{in}}\right)^{\gamma_2} \cdot (v_{in})^{\gamma_3}\right) \tag{5}$$

where the constants ($\alpha$, $\gamma_1$, $\gamma_2$, $\gamma_3$) are equal to (2.39, 0, 0.83, 2.88), (2.35, 0.045, 0.58, 3), and (1.7, 0.12, 0.48, 3) for the used PGSE, OGSE N=1 and N=2 sequences, respectively. Please refer to **Supplemental Materials** for more details. On the other hand, the extracellular space is regarded as a narrow interstitial space, and we approximate that: $k_{ex}^m \approx k_{ex} = k_{in} v_{in}/v_{ex}$.

2.1.3   Calculate $ADC_r$ under arbitrary gradient waveforms

The analytical expressions of the dMRI signal $S_r$ under restricted diffusion have been reported previously for PGSE and OGSE sequences(46), thereby enabling the calculation of $ADC_r$ directly. It is desirable to modify gradient waveforms in dMRI measurements for flexibility, but it is tedious and inefficient to derive such analytical expressions for complex gradient waveforms. Therefore, we develop a generalized framework based on the discretization of gradient waveforms $G(t)$ to calculate restricted signals and the corresponding $ADC_r$. As shown in Figure 1(c), a finite-duration, arbitrary gradient waveform can be discretized into a series of short pulses with a duration τ, then the computation of $S_r$ can be converted into a simple summation of matrix elements. Similarly, the $b$-value can be obtained based on this computational framework. The mathematical details are shown in the **Supplemental Materials**. The corresponding $ADC_r$ in Eq. (4) can be calculated by $ADC_r = -\log(S_r)/b$. The above generalized framework can be applied to arbitrary gradient waveforms, improving the adaptability of our microstructure imaging and the corresponding signal analysis method.

## 2.2 Data fitting

In this work, the solution of the microstructural parameters was transformed into a constrained optimization problem, which can be resolved by non-linear iterative methods. We assumed that the frequency-dependence of the extracellular diffusivity can be neglected for the acquisition protocols with the OGSE frequency ($f$) <100 Hz, and then there are four free parameters: $v_{in}$, $d$, $k_{in}$ ($k_{in} = 1/\tau_{in}$), and $D_{ex}$. If $f \geq 100$ Hz, it is necessary to introduce an additional free parameter $\beta$, and then we have $D_{ex} = D_{ex0} + \beta \cdot f$. 3D tumor cell density $\rho$ can then be calculated as $\rho = \frac{6v_{in}}{\pi d^3}$, and 2D cellularity = $2 \times \left(\frac{3V_{in}}{2\pi}\right)^{\frac{2}{3}}/d^2$ based on a tightly packed spherical cell model (47).

Another two methods, i.e., the aforementioned IMPULSED(13,48) and JOINT(36), are compared with the proposed one by fitting the same dMRI signals. IMPULSED cannot provide the exchange rate constant due to its ignorance of transcytolemmal water exchange. JOINT has incorporated the impact of water exchange into the IMPULSED framework, but its oversimplified assumptions still result in biases in estimating microstructural parameters(36).

All computations were performed in MATLAB R2017a (MathWorks, Natick, Massachusetts, USA) running on a 64-bit Linux machine with an Intel Core i7 3.6 GHz CPU.

## 2.3 Numerical simulations

A finite difference method(49) was used to simulate dMRI signals acquired from the sequences with the parameters shown in **Table 1(a)**. The tissue was modeled as tightly packed spherical cells (50). Here, the relaxation times are assumed to be spatially homogenous for simplicity (51). The used cell diameters $d$ were 8, 10, 12, 14, 16, and 18 μm, to represent typical cancer cell sizes. Meanwhile, the intracellular volume fractions $v_{in}$ were set to 42%, 51%, and 62%, and the water exchange rate constant $k_{in}$ were 0, 2.5, 5, 10, 14.2, and 20 s$^{-1}$, corresponding to $\tau_{in}$ values of ∞, 400, 200, 100, 70, and 50 ms. Furthermore, to evaluate the robustness of the proposed model, Rician noise was introduced into the simulated data.

## 2.4 Retrospective cell experiments in vitro

The details of the in vitro cell experiments have been reported previously(19). Briefly, fixed MEL cells were treated with different concentrations of saponin to form four groups, each with a different cell membrane permeability (highly related with $k_{in}$) while keeping other parameters the same. After removing the top liquid, the cell pellet samples were scanned using a Varian /Agilent 4.7 Tesla MRI Spectrometer (Palo Alto, California, USA) with a maintained temperature (17°C). The detailed acquisition parameters are summarized in **Table 1(b)**. The microstructural parameters were fitted by the biophysical models and then compared with the light microscopy-derived $d$, the constant gradient (CG) method-derived $k_{in}$, and the estimated reference values $v_{in}$.

## 2.5 Retrospective animal experiments in vivo

MDA-MB-231 tumors were formed in the right hind limb of mice. After tumor volumes reached around 50~100 mm$^3$ (week 0), 16 mice were treated with paclitaxel twice weekly at 20 mg/kg for three weeks (week 1, 2, 3). Another 11 mice did not receive any treatment and were included in the control group. The dMRI signal acquisitions were performed weekly for all included mice. During imaging, necessary measures were taken to anesthetize and immobilize the mice to ensure the high quality of the acquired images. Multiple axial slices covering the entire tumor of each mouse were acquired with a slice thickness of 2 mm. The matrix size was $32 \times 64$ with FOV= $16 \times 32$ mm, yielding an in-plane resolution of $0.5 \times 0.5$ mm. Other acquisition parameters are shown in **Table 1(c)**. Note that here the maximum OGSE frequency is 150 Hz (>100 Hz), so the frequency-dependent is non-negligible for the extracellular diffusivity. And there are five free parameters, $v_{in}$, $d$, $k_{in}$, $D_{ex0}$, and $\beta$, need to be fitted.

Due to the interindividual differences among mice, we focus on the variations of the tumor microstructural indicators during the treatment or no-treatment, and whether the treatment response can be accurately evaluated from these variations. We used EXCHANGE to extract voxel-wise microstructural information of tumors from the dMRI signals, and calculated the mean value ($\bar{x}$) and standard deviation (STD, $\sigma$) to summarize each parameter within the ROI. To quantify the weekly variations in the parameters, we introduced a dimensionless parameter $u$ as follows, inspired by the T-test in statistical analysis:

$$u = \frac{\bar{x}_i - \bar{x}_0}{\sqrt{\frac{\sigma_i^2}{N_i} + \frac{\sigma_0^2}{N_0}}} \tag{6}$$

where $(\bar{x}_0, \sigma_0, N_0)$ and $(\bar{x}_i, \sigma_i, N_i)$ are (mean value, STD, number of voxels) of the indicators obtained at week 0 and $i$ ($i = 1,2,3$), respectively. In addition, the tumor volume $V$ can be easily obtained by summing the number of voxels within the tumor ROI, which is also shown as a comparison with the EXCHANGE-derived microstructural parameters.

### 2.6 MRI of breast cancer patients with neoadjuvant chemotherapy

This human imaging study was approved by the medical ethics committee at Tsinghua University. Written informed consents were received from participants before inclusion. Other detailed information can be found in **Supplemental Materials**. In this work, we present the imaging results of two breast cancer patients with neoadjuvant chemotherapy.

The dMRI signal acquisitions were performed at weeks 0 (baseline), 6, and 12 after the start of chemotherapy, using a Siemens Prisma 3.0T scanner (Siemens Healthiness, Erlangen, Germany) with a 16-channel breast coil. Acquisition parameters are: TR/TE=5000/117ms; FOV=340 × 190 mm$^2$ (RL×AP); in-plane resolution = 2 × 2 mm; 10 slices; slice thickness = 4mm. Total scan time is ~6 minutes. Other diffusion sequence parameters are shown in **Table 1(d)**.

Please refer to the **Supplemental Materials** for additional details on numerical simulations, in vitro cell experiments, in vivo animal experiments, and human imaging studies.

## 3  RESULTS

### 3.1  Numerical simulations

Figure 2 shows the heat maps related to the error of the fitted parameters (red for overestimation and blue for underestimation), where each subfigure contains the results obtained from 36 different sets of simulations (6 $k_{in}$ values and 6 $d$ values) and the volume fraction $v_{in}$ is equal to 51% here. The signal-to-noise (SNR) was set to 45, obtainable in human breast cancer imaging(46). The fitted error for each parameter is calculated by the percentage deviation of the fitted value from the ground truth, where the "fitted value" is the mean of the fitted results in 100 trials after introducing random Rician noise, and the "ground truth" is the parameter set in the simulation. Note that for the cases of $k_{in} = 0 \text{ s}^{-1}$, we used $k_{in} = 2.5 \text{ s}^{-1}$ as the denominator to avoid non-convergent calculations.

First, for the intracellular volume fraction $v_{in}$, as reported previously, IMPULSED significantly underestimates $v_{in}$ (even up to -80%), and such bias increases rapidly with faster water exchange, i.e., larger $k_{in}$. By incorporating water exchange into biophysical modeling, although JOINT can improve the accuracy of fitted $v_{in}$, the distinct underestimation remains (up to -50%). By contrast, EXCHANGE provides the most accurate results for $v_{in}$, with a maximum error of only around -10%, which is a considerable improvement compared to the other two models. Then, for the cell diameter $d$, the results obtained by IMPULSED agree well with the ground truth, with the fitted error less than ±5%, which is consistent with the previous reports(19). JOINT overestimates the cell diameter for the cases of $d \leq 14$ μm and $k_{in} \geq 5 \text{ s}^{-1}$ (up to +20%). EXCHANGE provides the accurate estimations of $d$ for $d \leq 14$ μm, but underestimates it for the cases of $d \geq 16$μm and $k_{in} \geq 5 \text{ s}^{-1}$ (up to -15%). Finally, for the water exchange rate constants $k_{in}$, JOINT usually overestimates it for $d \geq 14$μm (up to +100%) and underestimates it for $d = 8$μm (up to -95%). By contrast, the $k_{in}$ values fitted by EXCHANGE agree much better with the ground truth, especially for the cases of $k_{in} \geq 10\text{s}^{-1}$ (less than ±15%), which highlights its reliability. The results similar to Figure 2 are shown in the **Supplemental Materials** for the cases with $v_{in}$=42% and 62%.

## 3.2  Cell experiments in vitro

The MEL cell diameters were measured as $11.34 \pm 1.68$ μm based on the light microscopy and the analysis approach reported previously(19). The corresponding volume-weighted cell diameter was ~12μm. The constant gradient (CG) method-derived $\tau_{in}$ values were $161.8 \pm 9.4, 157.8 \pm 8.9, 106.6 \pm 4.3, 59.4 \pm 3.7$ ms in the four groups of the in vitro cell experiments. Due to the lack of a standard method to measure the intracellular volume fraction in the cell samples. Here we roughly estimated the reference values of $v_{in}$ as 50%, 48%, 47%, and 38% for the four groups. The relevant details are provided in the **Supplemental Materials**.

Figure 3 summarizes the results of the in vitro cell experiments. For the water exchange rate constant, JOINT overestimates $k_{in}$, and the results obtained by EXCHANGE match the reference value well. For the intracellular volume fraction, the $v_{in}$ values fitted by IMPULSED and JOINT are lower than the reference values, which is consistent with the results in the numerical simulations and previous reports(19). The EXCHANGE model only slightly underestimates $v_{in}$ and its fitted values are in better agreement with the reference values. Finally, for the cell diameter, the fitted $d$ values from the three microstructural models are all in a reasonable range. This indicates that transcytolemmal water exchange has a limited impact on fitting cell sizes in dMRI-based microstructural imaging, as reported previously (19).

### 3.3 Animal experiments in vivo

Figure 4 shows the $u$-values of the five microstructural parameters ($k_{in}$, $v_{in}$, $d$, $D_{ex0}$, and $\beta$) and the percentage increase in tumor volume (compared to the original volume at week 0) for each surviving mouse at weeks 1, 2, and 3. For the control group, the percentage increased from week 1 to 3, indicating tumor growth. By contrast, this percentage within the treatment group increased from week 1 to 2 but subsequently decreased at week 3, implying tumor shrinkage and thus effective treatment. For the water exchange rate constant $k_{in}$, the average $u$ value within the treatment group, $\bar{u}_t$ (corresponds to the red pentagram, '$t$' for 'treatment'), was larger than 0 at weeks 1~3, which implies that the transcytolemmal water exchange was accelerated during the treatment compared with the initial state, i.e., the increased permeability as shown in Figure 1(d). On the other hand, all the $\bar{u}_c$ ($\bar{u}_c$ corresponds to the blue pentagram, '$c$' for 'control') values were less than 0. The above results indicate that the acceleration of water exchange measured in dMRI-based microstructural imaging may predict the response to tumor treatment. Then, for $v_{in}$ and $d$, all the $\bar{u}_t$ values except for $v_{in}$ at week 1 were less than 0, indicating that both the size and volume proportion of the tumor cells decreased during the treatment due to apoptosis (Figure 1(d)). By contrast, the $\bar{u}_c$ values were larger than or close to 0. The above findings suggest that the declines in the fitted $v_{in}$ and $d$ may also become potential biomarkers for the treatment response, although they are not as sensitive as the increase of $k_{in}$. Finally, for the two parameters characterizing extracellular diffusion, $D_{ex0}$ and $\beta$, due to the lack of theoretical modeling and relevant studies, our interpretation is very limited. All the $\bar{u}_c$ and $\bar{u}_t$ values for $D_{ex0}$ were close to 0, implying that this indicator is insensitive to both tumor growth and treatment. On the other hand, all the $\bar{u}_c$ values for $\beta$ were less than 0, while the corresponding $\bar{u}_t$ values were larger than 0. This suggests the potential of measuring variations in $\beta$ for treatment evaluation.

Furthermore, we also performed intergroup comparisons between the results of the treatment and control groups. The rank-sum test, also known as the Mann-Whitney U test, is used to evaluate the significance of the difference between the corresponding $u$ values of the two groups. As shown in Figure 4, both $V$ and $k_{in}$ demonstrated significant differences ($p < 0.01$) at weeks 2 and 3. However, the difference in the $u$ values for $k_{in}$ has already reached a significance level of $p < 0.01$ at week 1, while the difference in volume increase was not significant. This finding highlights the great potential of EXCHANGE in monitoring the early response of tumor treatment: that is, the EXCHANGE-derived micro-parameter $k_{in}$ shows earlier variations than the traditional macro-biomarker $V$. On the other hand, the intergroup differences for $v_{in}$, $d$, and $\beta$ reached a significance level of $p < 0.05$ at week 2 or 3, these parameters showed poorer sensitivity to tumor response compared with $k_{in}$.

As shown in Figure 5, we provide four sets of $k_{in}$ parametric maps (overlaid on the b0 image of the PGSE sequence) to illustrate the weekly variations of the EXCHANGE-derived $k_{in}$ on the tumor slices. In addition, the cell diameter is an intuitive and commonly used parameter in previous dMRI-based microstructural imaging, the corresponding $d$ maps are also shown for comparison. For the mice in the treatment group ($T\#1$ and $T\#2$), although the $k_{in}$ and $d$ maps are inhomogeneous, which reflects the heterogeneity within the tumor, the estimated $k_{in}$ increased overall from week 1, while the corresponding $d$ decreased, both of which are consistent with the cytomorphologic changes during treatment-induced apoptosis. On the other hand, for the mice in the control group without any treatment ($C\#1$ and $C\#2$), the variations of $k_{in}$ were unobvious, and only a slight decrease can be observed in the tumor slice of $C\#1$. By contrast, the increase of $d$ was relatively visible and the tumor volume (i.e., the area of the ROI) also became larger, which indicates that the tumors were still growing in these two mice.

### 3.4 MRI in breast cancer patients with neoadjuvant chemotherapy

We summarized the information on breast tumors and patients in **Supplemental Materials**. The SNR was ~40 in the tumors and ROIs were manually drawn by doctors (from Qinghai University Affiliated Hospital) based on PGSE images with $b = 1000 s/mm^2$. Figure 6 shows the EXCHANGE-derived microstructural parametric images at different time points (overlaid on the b0 image of PGSE), including $k_{in}$, $d$, and $v_{in}$. The rightmost column shows the statistical results (represented as the mean value and STD) of the fitted parameters within all ROI voxels. Although the variation in tumor volume is slight for Patient #1 and obvious for Patient #2, the neoadjuvant chemotherapy was effective for both patients according to the pathological conclusions from professional doctors. As shown in the figure, the water exchange rate constant $k_{in}$ was the most sensitive to the chemotherapy effect, and the fitted $k_{in}$ values within ROI increased overall during the treatment, with a significant difference of p<0.001 compared to the baseline results. On the other hand, the fitted volume fraction $v_{in}$ decreased during neoadjuvant chemotherapy, but in contrast to $k_{in}$, it showed a relatively lagged response to treatment for Patient #1 and the significance level of differences was slightly lower (p<0.01 vs. p<0.001). In addition, the fitted tumor cell diameter $d$ first decreased but then increased inversely, the significance level compared to the baseline results was also lower than that of $k_{in}$ (p<0.01 vs. p<0.001) for Patient #1. In summary, the above results from the two breast cancer patients with neoadjuvant chemotherapy are consistent with those of in vivo animal experiments, both demonstrating the clinical potential of EXCHANGE-derived microstructural parameters in monitoring tumor treatment response, especially the water exchange constant rate $k_{in}$.

## 4 DISCUSSION

Diffusion MRI (dMRI) is versatile and can provide rich information on biological tissue microstructures non-invasively. The dMRI-based cell size imaging and water exchange imaging are two branches of active dMRI research. The transcytolemmal water exchange is typically ignored in the dMRI-based cell size imaging to simply the quantitative biophysical models and cell size information can be estimated accurately(46). However, this simplification will lead to significant underestimation of the intracellular volume fraction, further resulting in estimation bias of cell density(19). In addition, the lack of estimation of water exchange rate constants also causes a loss of information on cell membrane permeability, which is another important indicator of tissue status(34). On the other hand, the dMRI-based water exchange imaging, such as FEXI(52), NEXI(25), SMEX(26), and diffusion-time-dependent diffusional kurtosis methods(27,28), usually assume exchangeable intra- and extracellular compartments but ignore the compartment sizes. They can obtain information on exchange rate constants and volume fractions but cannot provide compartmental size information. Moreover, these Kärger model-based methods neglect the impact of the restriction-induced edge-enhancement effect(45). The assumption of the short pulse approximation in the classic Kärger model greatly limits its applications in practical acquisitions, especially for the emerging OGSE sequence. Except for a few attempts based on this strong assumption(36,53), dMRI-based cell size imaging and water exchange imaging have been two parallel pathways in dMRI research with each ignoring the other one in their respective models. This is presumably due to the complexity of coupled water restriction and exchange, making it challenging to consider both phenomena in a single quantitative biophysical model.

In this work, we developed a new biophysical model to incorporate transcytolemmal water exchange into the dMRI-based cell size imaging, aiming to detect the microstructural features of tumors more comprehensively and accurately. Specifically, the transcytolemmal water exchange rate constant is an important indicator of cellular functions and it varies in tumor tissues undergoing treatment, as well as in disorders including Parkinson's(54) and Alzheimer's disease(55). Except for the Gd-based MR imaging(56,57), dMRI techniques have been widely used to characterize water exchange. The above mentioned FEXI method(58,59) employs double diffusion encoding imaging sequences(60) and provides an apparent exchange rate AXR=$1/(k_{in} \cdot v_{ex})$, where $k_{in}$ and $v_{ex}$ are coupled together, making it less specific to transcytolemmal water exchange (61). By contrast, the diffusion-time-dependent kurtosis imaging(27) improves the specificity of water exchange and reduces the SNR requirements, such as the CG method(62). However, these methods require independent measurements for water exchange, increasing the total scan time. Our proposed EXCHANGE method can simultaneously estimate cell diameter, intracellular volume fraction, and water exchange rate constant without the need for separate measurements. This effectively reduces the required scan time, which is desirable in clinical practice.

The validations by the numerical simulations and in vitro cell experiments show that EXCHANGE can extract more accurate and comprehensive microstructural information, especially for improving the underestimation of $v_{in}$ and providing an additional biophysical parameter $k_{in}$, as shown in Figure 2. The results of the retrospective in vivo animal experiments show the potential of the EXCHANGE-derived microstructural parameters as biomarkers for monitoring the response to tumor treatment. Especially for the water exchange rate constant $k_{in}$, its variations can reflect a significant difference between the tumors in the treatment and control groups earlier (week 1), compared to the tumor volume increase (week 2). Furthermore, the results on breast cancer patients with neoadjuvant chemotherapy also demonstrate the preliminary feasibility of the EXCHANGE method in monitoring the clinical effect of tumor treatment. As such, the EXCHANGE-based imaging method can be used to investigate the phenomena or processes accompanied by variations in the physical features of tumor tissues, such as treatment-induced cell apoptosis with increased cell membrane permeability and decreased cell diameter or volume proportion.

However, more validations through in vivo animal and human experiments are necessary to support the clinical feasibility of the proposed EXCHANGE method. In particular, dMRI data obtained from the joint acquisitions of PGSE and OGSE sequences need to be collected from more cancer patients undergoing chemotherapy, radiotherapy, or immunotherapy (14). Based on this, we can systematically investigate the variations of the EXCHANGE-derived microstructural parameters during the clinical treatment, and then evaluate the practical efficacy of their variations in monitoring the treatment response. Furthermore, more comprehensive multi-variate statistical methods also need to be implemented to analyze the experimental results, to make more accurate judgments on the treatment efficacy based on multiple indicators ($k_{in}$, $v_{in}$, $d$, ADC, and $V$) rather than only one. The above studies are beyond the scope of this paper and will be shown in our future work. Finally, some improvements to the current experimental design are necessary. Specifically, the results in the retrospective animal experiments (Figure 4) show a lower degree of differentiation between the treatment and control groups at week 3 compared to week 2, which is counterintuitive. We speculate that this abnormal finding is caused by inadequate experimental design. For the mice in the treatment group, if the treatment showed a significant effect at week 1 or 2, the mouse was usually sacrificed to confirm whether its tumor was cured. This implies that the mice still alive at week 3 might be relatively insensitive to the treatment. Similarly, for the control group, the mice that died at week 1 or 2 had faster tumor growth, while the tumors in the mice that survived at week 3 grew relatively slowly. Both the above reasons probably explain the lower degree of differentiation between the experimental results of the two groups at week 3. In future work, we will revise these shortcomings and improve the experimental design to obtain more accurate and solid results. In addition, it is interesting that the fitted tumor cell diameters went down at the early time point of neoadjuvant chemotherapy due to treatment-induced apoptosis but went up again at a later time point, as shown in Figure 6. It is unclear if it is caused by remaining non-responder cancer cells. More dMRI data at more time points and pathology results are needed in future studies to elucidate this point.

Another major limitation of this work is the lack of further investigation on tumor heterogeneity in the in vivo animal and human experiments. Although the use of mean values and STDs to highly summarize the EXCHANGE-derived parameters within the ROIs facilitates the intergroup analysis and comparisons, as shown in the results of the retrospective animal experiments, this analysis approach neglects the inter-voxel inhomogeneous distribution of imaging indicators and heterogeneity within tumors, which could contain useful information on local status and features. Therefore, more refined analysis methods, such as radiomics(63-65), need to be implemented to quantitatively investigate the heterogeneity of tumor tissues and then extract richer and more accurate information from the microstructural imaging results.

Here we revisit the extended Kärger model in Eq. (1). The classic Kärger model usually assumes the short pulse approximation and ignores water exchange during non-zero diffusion gradients, which is particularly inappropriate for the sequences with long gradient durations like OGSE. As shown in Figure 1 (c), a gradient waveform with a finite duration can be discretized into a series of short pulses, i.e., multiple propagators (66), and we can formulate the paired magnetization exchange equations, similar to Eq. (1), during each short pulse. Then the final dMRI signal can be obtained by solving the equations through all short pulses sequentially. This discretization-based computational approach has been approved to provide high accuracy in both Monte Carlo(67) and Finite Difference(49) simulations. However, the above numerical computation makes it impossible to derive an overall analytical expression of the dMRI signal, hindering the subsequent model fitting especially with the commonly used non-linear fitting methods. Therefore, we still insist on using only one set of equations to summarize the evolution of the intra- and extracellular magnetizations during the entire diffusion weighting and extend the classic Kärger model to Eq. (1) by approximating $\gamma^2 g^2 \delta^2 \approx b/T$. Although this strategy seems rough compared to the fine numerical computations, it enables us to derive an analytical expression of the dMRI signal, which effectively improves the solvability of the EXCHANGE method.

It is well known that dMRI measurements are dependent on gradient performance. Particularly, the OGSE sequences used in the current work expand the diffusion time range to provide better estimations of microstructural parameters, which requires higher maximum gradient strength $G_{max}$ (42). As shown in Table 1, the maximum b-values used in the human breast cancer imaging are lower compared to the other presented studies, due to the limitation of $G_{max}$ in clinical acquisitions. In the numerical simulations and cell experiments, we limited $G_{max} < 360 \, \text{mT/m}$ to demonstrate the feasibility of EXCHANGE, but this is not optimized for typical human imaging. We have previously demonstrated the feasibility of dMRI microstructural imaging in human breasts(46,68) and livers(69) with single-axis $G_{max} \leq 60 \, \text{mT/m}$. Other groups have shown $G_{max} \leq 45 \, \text{mT/m}$ is sufficient for time-dependent diffusion microstructural imaging in prostate cancer(70). With the recent rapid development of high-performance gradient coils with $G_{max} = 200 \, \text{mT/m}$ (71,72) and Human Connectome gradient coil version 1 and 2 with $G_{max} = 300 \, \text{mT/m}$ (73) and $500 \, \text{mT/m}$ (42), respectively. We expect no major obstacles to the application of EXCHANGE in human imaging, especially, $G_{max}$ is only $\sim 70 \, \text{mT/m}$ in our human breast cancer imaging. However, more optimization is desirable to further reduce the requirements of $G_{max}$.

## 5 CONCLUSION

In this work, we proposed a quantitative dMRI-based microstructural imaging, dubbed EXCHANGE, for the comprehensive characterization of cell size, density, and transcytolemmal water exchange in tumors. Validations in both numerical simulations and in vitro cell experiments showed promising results for accurate and robust estimation of microstructural parameters, including cell diameter, intracellular volume fraction, and water exchange rate constant. The results from in vivo animal experiments show the potential of EXCHANGE for monitoring the response of tumor treatment, and human breast cancer imaging demonstrates its preliminary clinical feasibility. This method provides new insights into characterizing tumor microstructural properties at the cellular level, suggesting a new, unique means to monitor tumor treatment response in clinical practice.

## ACKNOWLEDGMENTS

This work was funded by NIH grants R01 CA109106 (J.X.), R01 CA269620 (J.X.), R01 DK135950 (X.J.), and R21 CA270731 (J.X.).

**TABLES**

Table 1 Summary information on diffusion MRI sequences and acquisition parameters
**(a)**. In the numerical simulations.

| Sequence | $\delta/\Delta$ (ms) | $t_r/t_p$ (ms) | $f$ (Hz) | $b$ values (s/mm$^2$) |
|---|---|---|---|---|
| OGSE T-cos N=2 | 40.9/51.4 | 0.9/3.65 | 50 | 0, 200, 400, 600, 800, 1000, 1200, 1400, 1600, 1800, 2000 |
| OGSE T-cos N=1 | 40.9/51.4 | 0.9/8.65 | 25 | 0, 200, 400, 600, 800, 1000, 1200, 1400, 1600, 1800, 2000 |
| PGSE trapezoid | 12/74 | 0.9/10.2 | N/A | 0, 200, 400, 600, 800, 1000, 1200, 1400, 1600, 1800, 2000 |

Notes: T-cos means trapezoid-cosine waveform. N means the number of oscillating cycles, $\delta$ is the duration of the diffusion gradient, $\Delta$ means the separation time between two diffusion gradients, $t_r$ and $t_p$ are gradient rise time and plateau times, respectively.

**(b)** In the retrospective cell experiments

| Sequence | $\delta/\Delta$ (ms) | $f$ (Hz) | $b$ values (s/mm$^2$) |
|---|---|---|---|
| OGSE cos N=2 | 25/30 | 80 | 0, 250, 500, 750, 1000, 1250, 1500, 1750, 2000 |
| OGSE cos N=1 | 25/30 | 40 | 0, 250, 500, 750, 1000, 1250, 1500, 1750, 2000 |
| PGSE | 4/52 | N/A | 0, 250, 500, 750, 1000, 1250, 1500, 1750, 2000 |

Notes: Cos means cosine-shaped waveform. N means the number of oscillating cycles. $\delta$ is the duration of the diffusion gradient while $\Delta$ means the separation time between two diffusion gradients.

**(c)** In the retrospective animal experiment

| Sequence | $\delta/\Delta$ (ms) | $f$ (Hz) | $b$ values (s/mm$^2$) |
|---|---|---|---|
| OGSE cos N=3 | 20/25 | 150 | 0, 150, 300, 450, 600 |
| OGSE cos N=1 | 20/25 | 50 | 0, 375, 750, 1125, 1500 |
| PGSE | 3/46 | N/A | 0, 750, 1500, 2250, 3000 |

**(d)** In human breast cancer imaging study

| Sequence | $\delta/\Delta$ (ms) | $t_r/t_p$ (ms) | $f$ (Hz) | $b$ values (s/mm$^2$) |
|---|---|---|---|---|
| OGSE T-cos N=2 | 41.5/50 | 1.5/2.75 | 50 | 0, 250, 500 |
| OGSE T-cos N=1 | 41.5/60 | 1.5/7.75 | 25 | 0, 250, 500, 750, 1000 |
| PGSE trapezoid | 12/74 | 1.5/9 | N/A | 0, 250, 500, 750, 1000 |

**FIGURES**

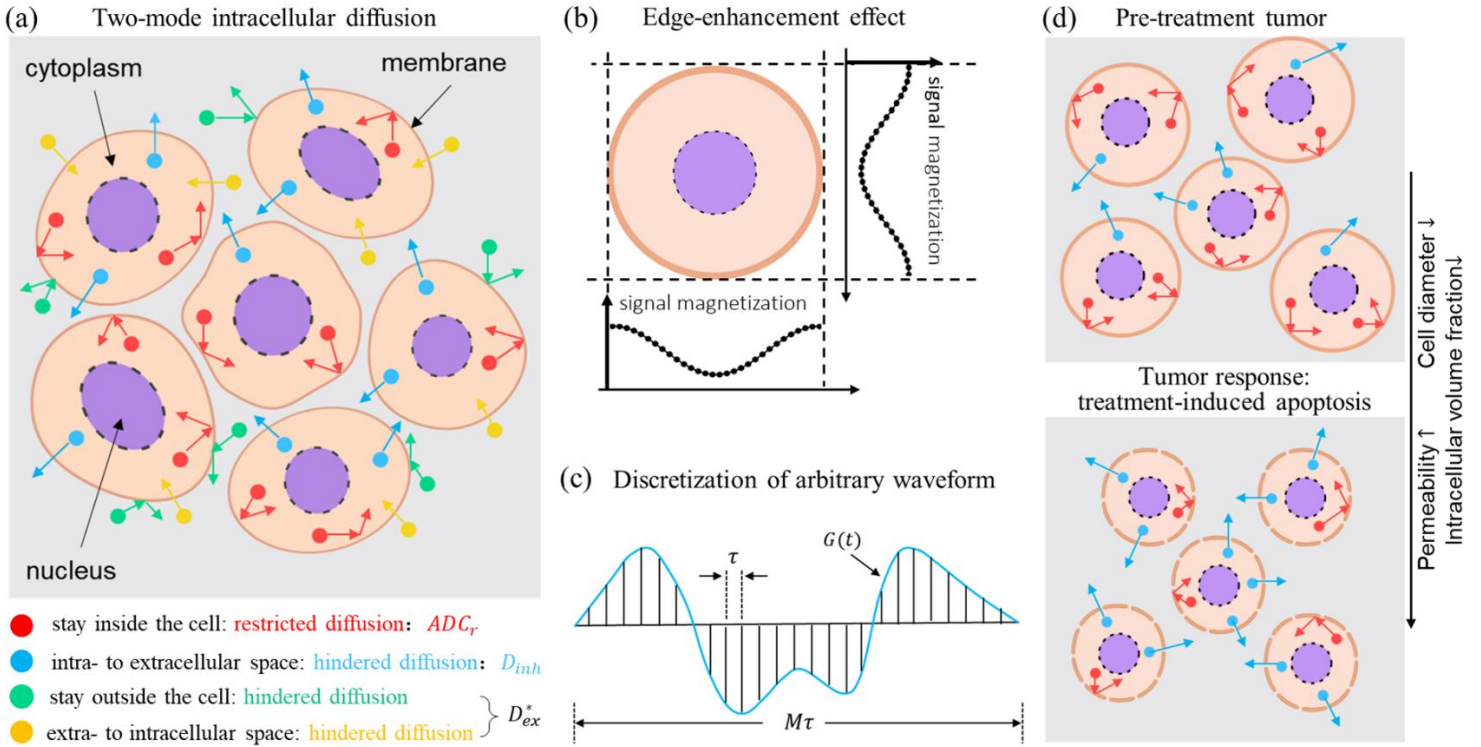

Fig. 1 Schematic diagrams of the main studies in this work. (a). four different diffusion movements of water molecules in the intra- and extracellular spaces. (b). restricted-induce edge-enhancement effect. (c). the discretization of an arbitrary diffusion gradient waveform $G(t)$. $\tau$ is the time duration of each short pulse. N is the total number of short pulses. (d). treatment-induced microstructural variations in tumor tissues.

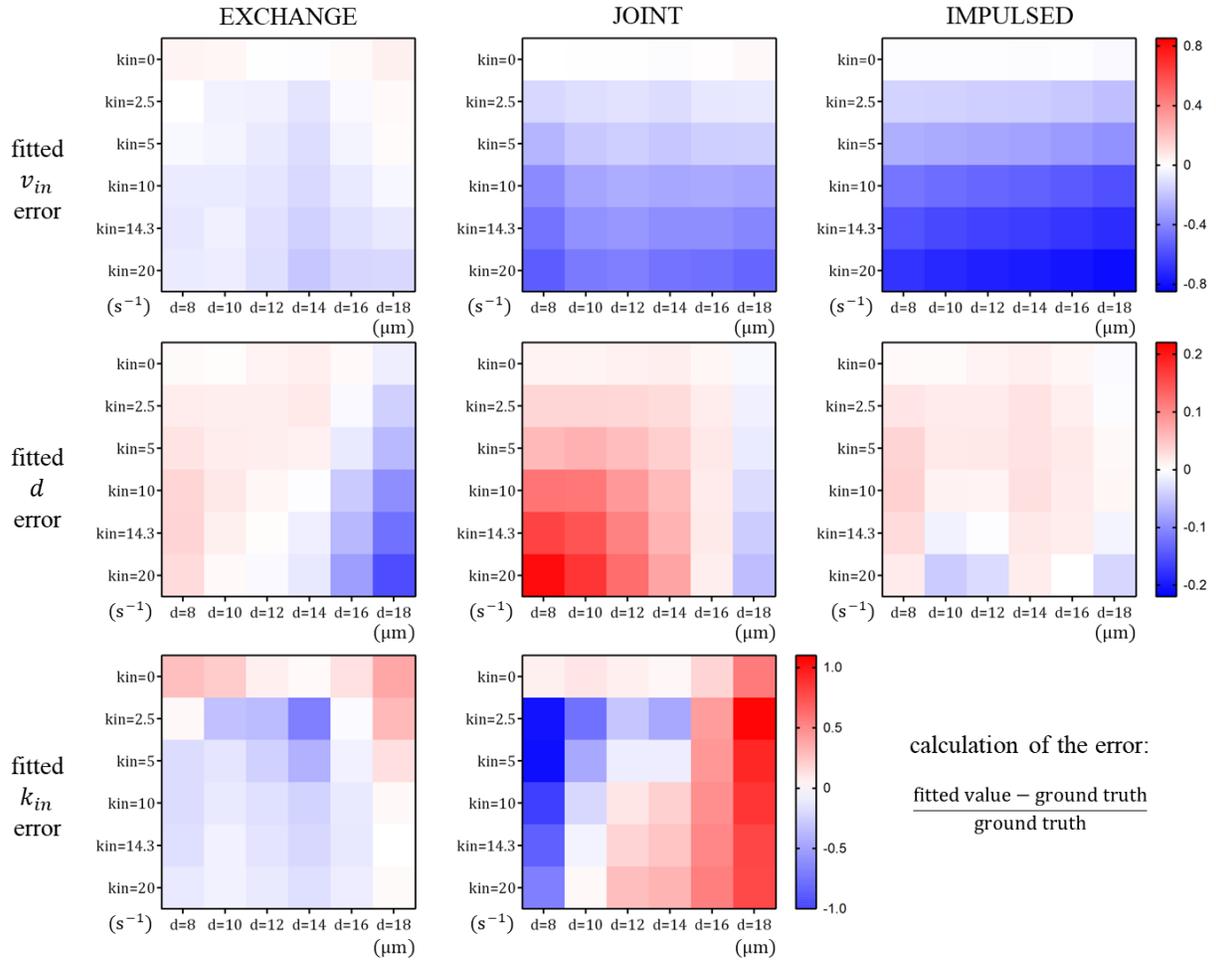

Fig. 2. The heat maps related to the fitted error of three different DW-MRI-based microstructural imaging Methods (EXCHANGE, JOINT, and IMPULSED) on the numerical simulations.

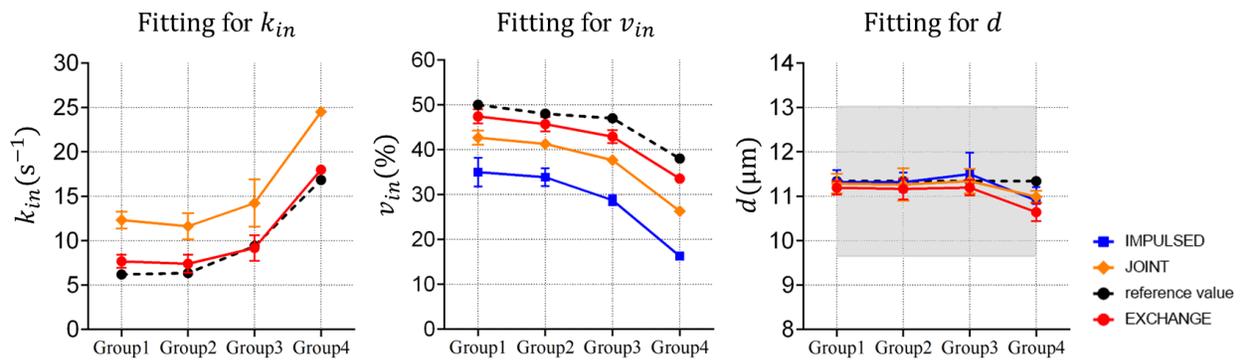

Fig. 3. Results from the IMPULSED, JOINT, and EXCHANGE models in the four groups of retrospective cell experiments in vitro. The shaded area indicates the STD of light-microscopy obtained cell sizes.

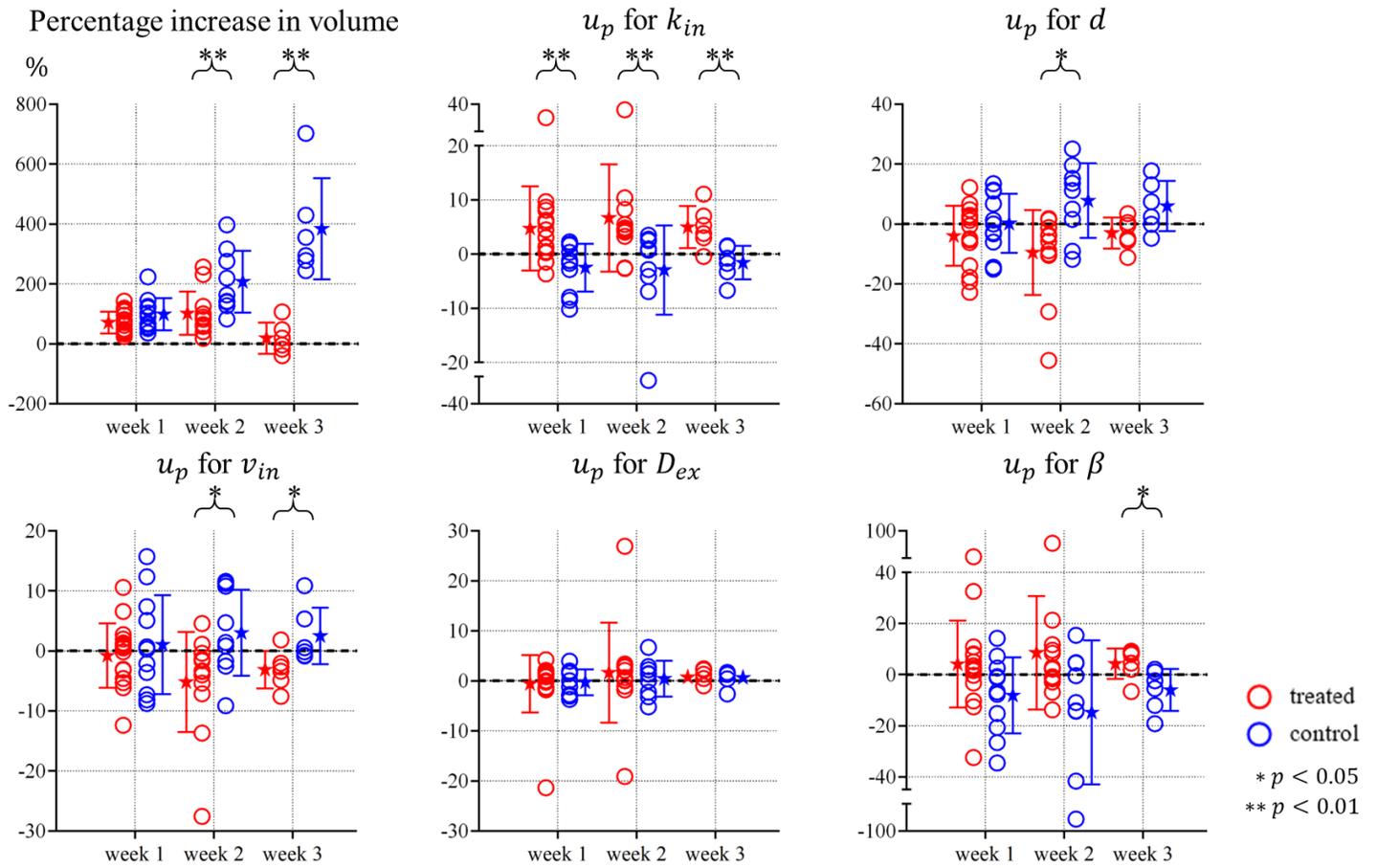

Fig. 4 The results for the intergroup comparisons of variations in tumor volume and individual microstructural parameters in retrospective animal experiments in vivo.

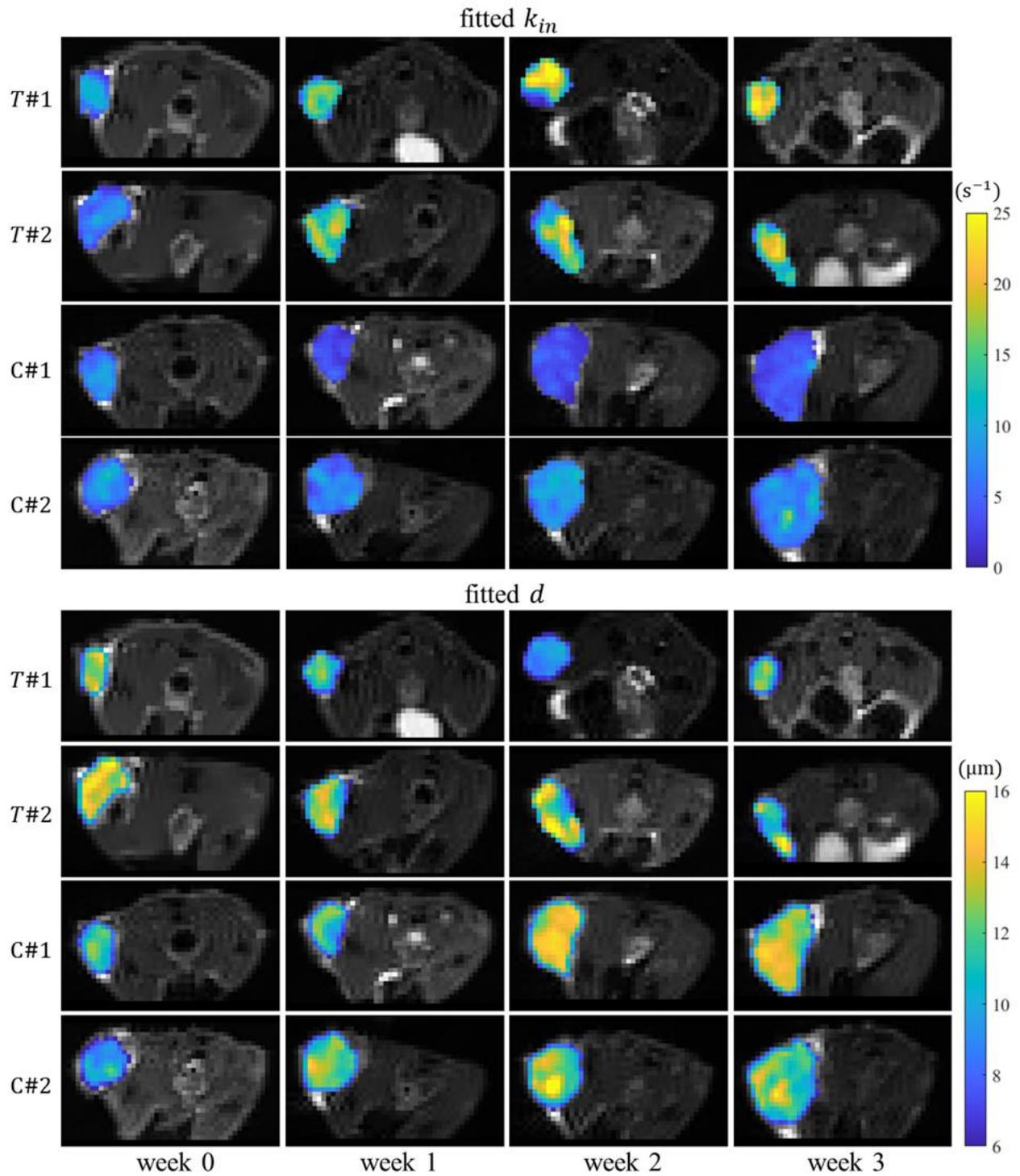

Fig. 5. EXCHANGE-derived $k_{in}$ and $d$ mappings of the members in the treatment ($T\#1$ and $T\#2$) and control groups ($C\#1$ and $C\#2$), which are overlaid on the b0 image of the PGSE sequence.

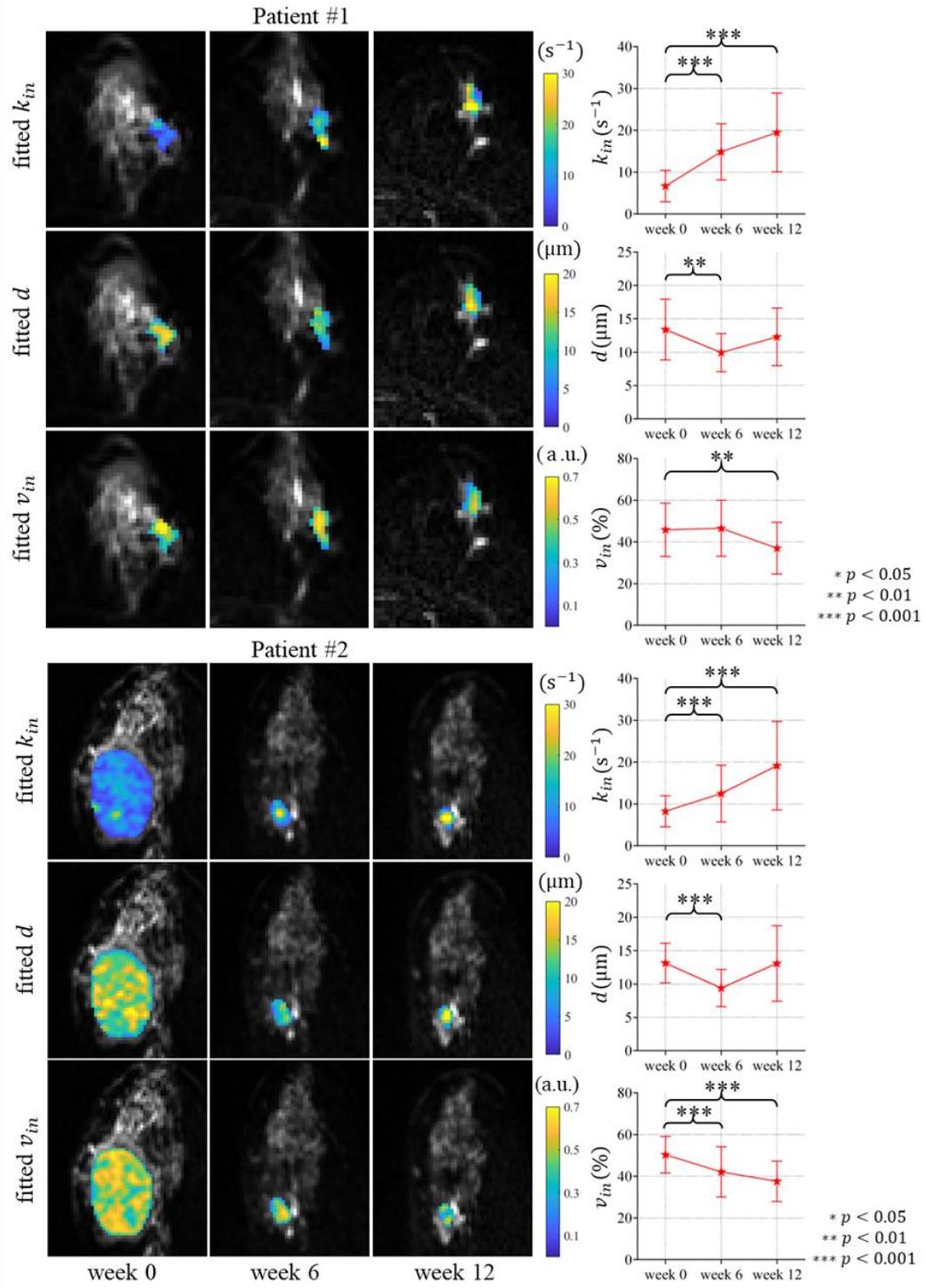

Fig. 6. Left columns: EXCHANGE-derived microstructural parameter mappings (overlaid on the b0 image of PGSE), including $k_{in}$, $d$, and $v_{in}$. Rightmost column: the statistical results of the fitted parameters within all ROI voxels.

# SUPPLEMENTAL MATERIALS

**Analytical expression of the proposed biophysical model**

The solution of the differential Eq. (1) is:

$$S = V_1 \exp(-bD_1^*) + (1 - V_1) \exp(-bD_2^*) \qquad (S1)$$

where the terms $D_1^*$, $D_2^*$, and $V_1$ are:

$$D_1^* = \frac{A_{in} - A_{ex} - D_Q}{2}$$

$$D_2^* = \frac{A_{in} - A_{ex} + D_Q}{2} \qquad (S2)$$

$$V_1 = 1 - \frac{(A_{ex} - A_{in} + D_Q - 2Tk_{ex}^m/b)v_{ex} + (A_{in} - A_{ex} + D_Q - 2Tk_{in}^m/b)v_{in}}{2D_Q}$$

$v_{in}$ and $v_{ex}$ are the volume fractions of the intra- and extracellular compartments respectively, and $v_{in} + v_{ex} = 1$. The other unknown parameters are computed by:

$$A_{in} = D_{in}^* + Tk_{in}^m/b$$
$$A_{ex} = D_{ex}^* + Tk_{ex}^m/b \qquad (S3)$$
$$D_Q = \sqrt{(A_{in} - A_{ex})^2 + 4T^2 k_{in}^m k_{ex}^m / b^2}$$

Substituting Eqs. (S2) and (S3) into (S1), we can obtain the final analytical expression of the proposed biophysical model.

**Estimation of the probability of crossing the cell membrane**

First, some simplified assumptions were necessary: (1) water molecules are uniformly distributed in each tumor cell; (2) the probability of molecule diffusion is identical in all directions; (3) there is no collision between molecules during the movement.

Then we calculated the average displacement of intracellular water molecules moving from their current positions to the cell membrane. As shown in Figure S1(a), the displacement $s(r, \varphi)$ of a water molecule is given by:

$$s(r, \varphi) = \sqrt{R^2 - r^2 \sin^2 \varphi} - r\cos\varphi \qquad (S4)$$

Due to the central and rotational symmetry of the sphere, it is easy to calculate the average displacement $\bar{s}$ of all water molecules diffusing to the boundary:

$$\bar{s} = \frac{1}{V} \int_0^R 4\pi r^2 \int_0^\pi s(r, \varphi) \frac{\sin\varphi}{2} d\varphi \, dr = \frac{3}{4} R \qquad (S5)$$

According to the mean free path formula: $\bar{s} = \sqrt{2D_{in}\bar{t}}$, the corresponding transportation time $\bar{t}$ is:

$$\bar{t} = \frac{\bar{s}^2}{2D_{in}} = \frac{(3R/4)^2}{2D_{in}} \tag{S6}$$

For each intracellular molecule that reaches the boundary, it has a probability $p$ to cross the membrane. If not, it remains in the cell and diffuse to the membrane again, as shown in Figure S1 (b). Here, we assume that the molecule has the same probability of travelling from its current position (such as point A) to other positions (point B, C, D or E) on the cell membrane, then the corresponding average displacement $\bar{s}^*$ is:

$$\bar{s}^* = \int_0^\pi \left(2R\cos\frac{\varphi}{2}\right)\frac{\sin\varphi}{2} d\varphi = \frac{4}{3}R \tag{S7}$$

And the transportation time $\bar{t}^*$ is:

$$\bar{t}^* = \frac{\bar{s}^{*2}}{2D_{in}} = \frac{(4R/3)^2}{2D_{in}} \tag{S8}$$

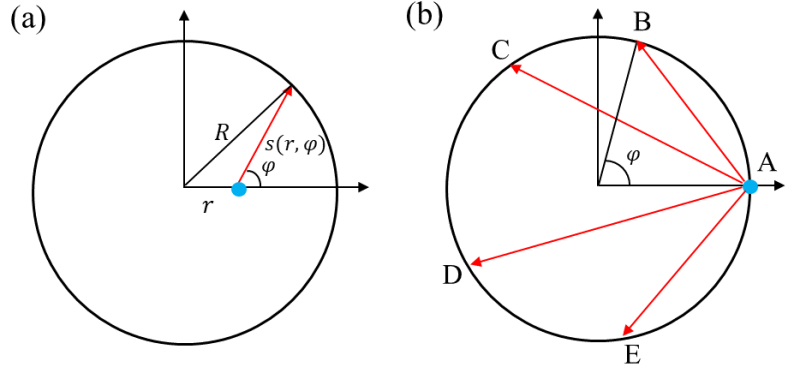

**Figure S1.** Schematic diagrams of the water molecule displacement calculation for two-mode intracellular diffusion.

When a molecule reaches the boundary again, the probability of crossing the cell membrane is still $p$. If it stays in the intracellular space, the average time taken to reach the cell membrane for the next time is still $\bar{t}^*$. As such, **Table S1** shows the probability of molecules crossing the membrane upon reaching the boundary for $i$-th time, along with the corresponding average diffusion time. This process is similar to a mathematical binomial distribution, then the expectation of the time that a water molecule remains in the intracellular space, i.e., $\tau_{in}$ $(= 1/k_{in})$, can be expressed as:

$$\tau_{in} = \sum_{i=1}^{\infty} p(1-p)^{i-1}(\bar{t} + (i-1)\bar{t}^*) = \bar{t} + \frac{1-p}{p}\bar{t}^* \tag{S9}$$

Therefore, the probability of crossing the membrane can be estimated by:

$$p = \frac{\bar{t}^*}{\bar{t}^* + \tau_{in} - \bar{t}} = \frac{\bar{t}^*}{\bar{t}^* + 1/k_{in} - \bar{t}} \tag{S10}$$

**Table S1** The probability and corresponding average diffusion time for water molecules crossing the membrane.

| Number of times a molecule reaches the membrane before crossing it | Probability | Average diffusion time |
|---|---|---|
| 1 | $p$ | $\bar{t}$ |
| 2 | $(1-p)p$ | $\bar{t} + \bar{t}^*$ |
| 3 | $(1-p)^2 p$ | $\bar{t} + 2\bar{t}^*$ |
| ... | ... | ... |
| $i$ | $(1-p)^{i-1}p$ | $\bar{t} + (i-1)\bar{t}^*$ |

**Correction for the edge-enhancement effect**

Based on the dimensional analysis, we construct a simple correction form for the magnetization exchange rate constant $k_{in}^m$ to eliminate the impact of restricted-induced edge-enhancement effect, as shown in Eq. (5) in the text. Then the unknown constants ($\alpha$, $\gamma_1$, $\gamma_2$, $\gamma_3$) can be determined from a subset of the numerical simulation data, where $d = 10, 14, 18$ μm, $v_{in} = 42\%, 51\%, 62\%$, and $\tau_{in} = 400, 100, 50$ ms, the other parameters were consistent with those in the text.

Specifically, we proposed the following optimization question and solved it by non-linear iterative fitting to obtain the optimal values of ($\alpha$, $\gamma_1$, $\gamma_2$, $\gamma_3$):

$$Y = \min_{\alpha, \gamma_1, \gamma_2, \gamma_3} \sum k_{in}^2 \cdot |\hat{S}(\alpha, \gamma_1, \gamma_2, \gamma_3) - S|^2 \tag{S11}$$

where $S$ is the dMRI signals (noise-free) generated by the finite difference simulation, and $\hat{S}$ is the signals obtained by our proposed EXHANGE model. We search for the optimal set of parameters ($\alpha$, $\gamma_1$, $\gamma_2$, $\gamma_3$) to minimize the overall difference between the $\hat{S}$ and $S$. Note that $k_{in}^2$ is used as the weight to improve the applicability of the model for faster water exchange (larger $k_{in}$ values).

**Calculate $S_r$ and *b*-value under arbitrary gradient waveforms**

Based on the velocity correlation function developed by Stepisnik, the restricted signal $S_r$ can be written as $S_r = \exp(-\varphi_r)$, where:

$$\varphi_r = \frac{\gamma^2}{2} \sum_k B_k \int_0^{TE} dt_1 \int_0^{TE} dt_2 \exp(-a_k D_{in}|t_1 - t_2|) G(t_1) G(t_2) \tag{S12}$$

where $B_k$ and $a_k$ are geometric parameters. After discretizing the gradient waveform, the exponential attenuation factor $\varphi_r$ can be expressed as:

$$\varphi_r = \frac{\gamma^2}{2} \sum_k B_k \sum_{i=1}^{M} \sum_{j=1}^{M} \int_{t_{i-1}}^{t_i} dt_1 \int_{t_{j-1}}^{t_j} dt_2 \exp(-a_k D_{in}|t_2 - t_1|) G(t_1) G(t_2) \tag{S13}$$

where $t_i = i\tau$. Then a series of $M \times M$ symmetric matrices $\mathbf{C}^k$ are defined, whose elements are:

$$C_{ij}^k = \int_{t_{i-1}}^{t_i} dt_1 \int_{t_{j-1}}^{t_j} dt_2 \exp(-a_k D_{in}|t_2 - t_1|) G(t_1) G(t_2) \tag{S14}$$

Each element $C_{ij}^k$ can be calculated by the following approach: first, approximate $G(t)$ during each short pulse as:

$$G(t) \approx G\left(t_{i-1} + \frac{\tau}{2}\right), \text{ for: } t_{i-1} < t < t_i \text{ (i.e. } t_{i-1} + \tau\text{)} \tag{S15}$$

Then for $i < j$, we have:

$$C_{ij}^k = C_{ji}^k = \int_{t_{i-1}}^{t_i} G(t_1) \exp(a_k D_{in} t_1) dt_1 \cdot \int_{t_{j-1}}^{t_j} G(t_2) \exp(-a_k D_{in} t_2) dt_1$$

$$= \frac{2(\cosh(a_k D_{in} \tau) - 1)}{(a_k D_{in})^2} G\left(t_{i-1} + \frac{\tau}{2}\right) G\left(t_{j-1} + \frac{\tau}{2}\right) \exp(-a_k D_{in}(t_j - t_i)) \tag{S16}$$

And for $i = j$:

$$C_{ii}^k = \int_{t_{i-1}}^{t_i} dt_1 \int_{t_{i-1}}^{t_i} dt_2 \exp(-a_k D_{in}|t_2 - t_1|) G(t_1) G(t_2)$$

$$= \frac{2G\left(t_{i-1} + \frac{\tau}{2}\right)^2}{(a_k D_{in})^2} (a_k D_{in}\tau + \exp(-a_k D_{in}\tau) - 1) \tag{S17}$$

Based on Eqs. (S16) and (S17), the attenuation factor $\varphi_r$ can be easily computed as:

$$\varphi_r = \frac{\gamma^2}{2}\sum_k B_k \sum_{i=1}^{M}\sum_{j=1}^{M} C_{ij}^k = \frac{\gamma^2}{2}\sum_k B_k \text{sum}(\mathbf{C}^k) \tag{S18}$$

where "sum" means to sum all matrix elements. Compared to the reported multiple-propagator method, the above expression is more concise in form, and the elements of $\mathbf{C}^k$ do not involve series expansions or summations, which greatly shorten the computation time.

In addition, the $b$-value can be expressed as:

$$b = \gamma^2 \int_0^{TE} \left(\int_0^t G(t')dt'\right)^2 dt \tag{S19}$$

We introduce an auxiliary function $f(t)$:

$$f(t) = \int_0^t G(t')dt' \tag{S20}$$

Then the $b$-value can be rewritten as:

$$b = \gamma^2 \int_0^{TE} f(t)^2 dt = \gamma^2 \sum_{i=1}^{M} \int_{t_{i-1}}^{t_i} f(t)^2 dt \approx \gamma^2 \sum_{i=1}^{M} \frac{\tau}{2}(f(t_{i-1})^2 + f(t_i)^2) \tag{S21}$$

where:

$$f(t_i) = f(i\tau) = \sum_{j=1}^{i} \int_{t_{j-1}}^{t_j} G(t')dt' \approx \sum_{j=1}^{i} \tau G\left(t_{i-1} + \frac{\tau}{2}\right) \tag{S22}$$

Based on Eq. (S22), the $b$-value can be computed by Eq. (S21) for arbitrary gradient waveforms.

**Details on numerical simulations**

Diffusion signals were simulated by a finite difference method, where the tumor tissue was modeled as tightly packed, spherical cells on a face-centered-cubic lattice. The intra- and extracellular diffusion coefficients without the influences of cell membranes were set to 1.56 and 2 μm²/ms, respectively. In addition to the basic information presented in the main text, note that here we used cell membrane permeability $P_m$ to reflect the rate of transcytolemmal water exchange, instead of $k_{in}$ or $\tau_{in}$. A larger $P_m$ indicates more frequent water exchange. Specifically, the relationship between $P_m$ and $k_{in}$ is:

$$\frac{1}{P_m} = \frac{6}{k_{in} \cdot d} - \frac{d}{10 D_{in}} \qquad (S23)$$

*Rician* noise was also introduced into the simulated signals to investigate the robustness of the proposed model:

$$S^* = \sqrt{(S + \eta_1)^2 + \eta_2^2} \qquad (S24)$$

where $S$ denotes the original signal generated by the finite difference method while $S^*$ is the signal with added noise, $\eta_1$ and $\eta_2 \sim N(0, \sigma^2)$, $\sigma$ is a parameter to control the level of noise. Practically, the data acquisition is usually repeated and geometric averaging is used to reduce the influence of noise. Here, in each trial, we generated three noise signals for each simulated signal and computed their geometric mean for data fitting. For each simulation, the above test with introducing *Rician* noise was repeated 100 times.

All data fittings were performed using non-linear iterative methods in MATLAB to generate microstructure parametric maps on a voxel-wise basis. The fitting parameter ranges were limited by reasonable values, in this part, $0 \le d \le 30$ μm (covering most tumor cell size), $0 \le v_{in} \le 1$, $0 \le k_{in} \le 100$ $S^{-1}$ (corresponding to $\tau_{in} > 10$ ms, covering known water exchange rate constants), and $0 \le D_{ex} \le 3$ μm²/ms (the free water diffusion coefficient is ~3 μm²/ms at 37°C). $D_{in}$ was fixed as 1.56 μm²/ms in the data analysis to stabilize fittings as reported previously.

**Details on in vitro cell experiments**

Murine erythroleukemia (MEL) cancer cells purchased from the American Type Culture Collection grew under standard culture conditions. Large-scale MEL cells were cultured in multiple 150-mm dishes; after being spun down and washed with PBS, the cells were fixed with 4% paraformaldehyde in PBS for around two hours. Then cells were washed and divided into four groups at a cell density of $4.2 \times 10^7$ cell/ml, each treated with 0, 0.01, 0.025, and 0.05% saponin at room temperature for 30 minutes to induce different degrees of cell membranes at low concentration, so other cell microstructure could be unchanged. For each saponin treatment concentration, the cells were evenly divided into six 0.65 ml microtubes. After centrifuging at 6000 g for 2 min, the top fluid was carefully removed and the cell pellet samples were then used for MR experiments. A small aliquot of the sample was spotted on glass slides and imaged directly under phase contrast microscopy to estimate the cell diameter.

All dMRI measurements were performed on a Varian 4.7 Tesla MRI spectrometer (Palo Alto,

California, USA). The sample temperature was maintained at ~17°C using a cooling water circulation system. In addition, the intracellular water exchange lifetime $\tau_{in}$ was estimated using constant gradient (CG) experiments with a stimulated echo (STEAM) sequence. Then the measured $\tau_{in}$ was converted to $k_{in}$ by $k_{in} = 1/\tau_{in}$.

The constraints on the free parameters in this part are: $0 \leq d \leq 30$ μm, $0 \leq v_{in} \leq 1$, $0 \leq k_{in} \leq 100$ S$^{-1}$, and $0 \leq D_{ex} \leq 3$ μm²/ms. Note that $D_{in}$ is usually temperature-dependent: $D_{in} \sim T^{\frac{3}{2}}$, and therefore the $D_{in}$ value was fixed as $1.56 \cdot \left(\frac{273+17}{273+37}\right)^{\frac{3}{2}} \approx 1.4$ μm²/ms in the cell experiments.

**Details on in vivo animal experiments**

The MDA-MB-231 tumor was formed in the right hind limb of each experimental mouse. For high-quality imaging, mice were anesthetized with a 2%/98% isoflurane/oxygen mixture before and during scanning, and they were restrained in a customized Teflon animal holder with a tooth bar and a head bar during imaging. All images were acquired with a Varian DirectDriveTM horizontal 4.7 T magnet (Varian Inc., Palo Alto, CA). The magnet bore temperature was kept at ~37°C using a warm-air feedback system, and $D_{in}$ was fixed as 1.56 μm²/ms in the data analysis.

It should be noted that there is an additional free parameter $\beta$ in the model fitting of this part because the used maximum frequency exceeds 100 Hz, and we limited its range to $0 \leq \beta \leq 10$ μm². The constraints on the other free parameters are: $0 \leq d \leq 30$ μm, $0 \leq v_{in} \leq 0.7$ (the space utilization rate of densest packing is $\pi/3\sqrt{2} \approx 0.7$), $0 \leq k_{in} \leq 30$ S$^{-1}$, and $0.8 \leq D_{ex} \leq 2$ μm²/ms. Whether in the animal experiments or human imaging below, we have tightened the constraints on some free parameters in model fitting to obtain better results.

**Details on human imaging**

The detailed information on breast cancer patients and tumors are shown in the following table

Table S2 Information on breast cancer patients and their tumor types

| Patient | age | Pathological diagnosis | Clinical stage | Ki-67 | ER | PR | HER2 |
|---|---|---|---|---|---|---|---|
| 1 | 37 | Invasive ductal carcinoma left breast | IIB | 30% | – | – | – |

| | | Invasive ductal | | | | | |
|---|---|---|---|---|---|---|---|
| 2 | 41 | carcinoma right breast | IIIA | 70% | – | – | – |

The constraints on the free parameters in human imaging are: $0 \leq d \leq 30$ µm, $0 \leq v_{in} \leq 0.7$, $0 \leq k_{in} \leq 50\ \text{S}^{-1}$, and $0.5 \leq D_{ex} \leq 3\ \text{µm}^2/\text{ms}$. the $D_{in}$ value was also fixed as $1.56\ \text{µm}^2/\text{ms}$.

**Estimation of $v_{in}$ for in vitro cell samples**

With a lack of standard methods, the ground-truth $v_{in}$ of cell culture samples was not estimated in the cell experiments. However, based on the fitting results from the JOINT model in the numerical simulations, and combined with the measured values of $d$ and $\tau_{in}$, we can provide a rough estimation of the reference values for $v_{in}$ by the linear interpolation. Specifically, the true intracellular volume fraction is denoted as $v_{in}^{\text{true}}$, and the corresponding fitted value of the JOINT model is denoted as $v_{in}^{\text{JOINT}}$. In the four groups of cell experiments, $d \approx 12$ µm. For the 1st and 2nd groups, $\tau_{in} \approx 160$ ms and $k_{in} \approx 6.25\ \text{s}^{-1}$. From the results shown in Figure 4: $v_{in}^{\text{JOINT}} \approx 36\%$ for $v_{in}^{\text{true}} = 42\%$, and $v_{in}^{\text{JOINT}} \approx 43\%$ for $v_{in}^{\text{true}} = 51\%$. Then the correspondence between $v_{in}^{\text{true}}$ and $v_{in}^{\text{JOINT}}$ can be approximated as:

$$v_{in}^{\text{true}} \approx 1.29 v_{in}^{\text{JOINT}} - 4.29\% \tag{S25}$$

Substitute the fitting results of $v_{in}^{\text{JOINT}}$ in the 1st and 2nd groups into the above equation, and the corresponding $v_{in}^{\text{true}}$ values are 50% and 48%, respectively.

In the same way, for the other two groups of cell experiments, $v_{in}^{\text{true}}$ can also be obtained by the linear interpolation method. For the 3rd group: $\tau_{in} \approx 107$ ms and $k_{in} \approx 9.35\ \text{s}^{-1}$, then $v_{in}^{\text{JOINT}} \approx 33\%$ for $v_{in}^{\text{true}} = 42\%$, and $v_{in}^{\text{JOINT}} \approx 40\%$ for $v_{in}^{\text{true}} = 51\%$, thus the correspondence is:

$$v_{in}^{\text{true}} \approx 1.29 v_{in}^{\text{JOINT}} - 0.43\% \tag{S26}$$

Substitute $v_{in}^{\text{JOINT}}$ in the 3rd group into Eq. (S26) and obtain $v_{in}^{\text{true}} \approx 47\%$. Finally, for the 4th group: $\tau_{in} \approx 60$ ms and $k_{in} \approx 16.67\ \text{s}^{-1}$, then $v_{in}^{\text{JOINT}} \approx 29\%$ for $v_{in}^{\text{true}} = 42\%$, and

$v_{in}^{JOINT} \approx 34\%$ for $v_{in}^{true} = 51\%$, thus the correspondence is:

$$v_{in}^{true} \approx 1.8 v_{in}^{JOINT} - 10.2\% \tag{S27}$$

Substitute $v_{in}^{JOINT}$ in the 4th group into Eq. (S27) and obtain $v_{in}^{true} \approx 38\%$. Considering that the above results were not measured in the experiment, we refer to them (50%, 48%, 47%, 38%) as reference values of the intracellular volume fraction.